# A Novel Attention-Augmented Wavelet YOLO System for Real-time Brain Vessel Segmentation on Transcranial Color-coded Doppler

Wenxuan Zhang#, Shuai Li#, Xinyi Wang, Yu Sun, Hongyu Kang, Pui Yuk Chryste Wan, Jing Qin, Yuanpeng Zhang, Yong-Ping Zheng*, and Sai-Kit Lam*

***Abstract*—The Circle of Willis (CoW), vital for ensuring consistent blood flow to the brain, is closely linked to ischemic stroke. Accurate assessment of CoW is important for identifying individuals at stroke risk and guiding appropriate clinical management. Among existing imaging methods, Transcranial Color-coded Doppler (TCCD) exhibited its unique superiority for CoW assessment owing to its radiation-free nature, affordability, and accessibility. Yet, reliable TCCD assessments rely heavily on the expertise of operators to identify anatomical landmarks and execute accurate angle correction, greatly restricting the adoption of TCCD for CoW assessments. There is a pressing need for an AI-powered real-time CoW auto-segmentation system that can efficiently capture cerebral arteries. Such a system is currently unavailable, and no prior studies have explored AI-driven cerebrovascular segmentation using TCCD. Herein, we proposed a novel Attention-Augmented Wavelet YOLO (AAW-YOLO) network tailored for TCCD data, in the hope of offering real-time guidance of brain vessel auto-segmentation in CoW. TCCD data were prospectively collected with 738 annotated frames and 3,419 labeled artery instances to establish a high-quality dataset for model training and evaluation. The proposed AAW-YOLO demonstrated its superiority in segmenting both ipsilateral and contralateral CoW vessels (average Dice (0.901), IoU (0.823), Precision (0.882), Recall (0.926), and mAP (0.953)) with high computational efficiency (inference speed: 14.199 milliseconds per frame). It offers a practical solution to eliminate the reliance on operators' experience for TCCD cerebrovascular screening, with potential applications in routine clinical workflows and resource-constrained settings. Future research will explore bilateral modeling and larger-scale validation.**

This project was supported by the Hong Kong Polytechnic University (University Grants Committee) (P0053754, P0043005) and The Innovation and Technology Fund of Hong Kong Government (MRP/022/18X). # Wenxuan Zhang and Shuai Li contributed equally.

Wenxuan Zhang, Shuai Li, Xinyi Wang, Yu Sun, Kang Hongyu, and Chryste Pui Yuk Wan are with the Department of Biomedical Engineering, The Hong Kong Polytechnic University, Hong Kong SAR, China.

Jing Qin is with the Centre for Smart Health, School of Nursing, The Hong Kong Polytechnic University, Hong Kong SAR, China

Yuanpeng Zhang is with the Department of Medical Informatics, Nantong University, Nantong, China

Yong-Ping Zheng and Sai-Kit Lam are with the Department of Biomedical Engineering, The Hong Kong Polytechnic University, Hong Kong SAR, China, and also with the Research Institute for Smart Ageing, The Hong Kong Polytechnic University, Hong Kong SAR, China.

Co-corresponding authors: Yong-Ping Zheng (e-mail: yongping.zheng@polyu.edu.hk; Tel.: +852-2766-7664); Sai-Kit Lam (e-mail: saikit.lam@polyu.edu.hk; Tel.: +852-2766-4731).

## I. Introduction

STROKE is a crippling disease worldwide, ranking the second leading cause of mortality and a major contributor to long-term disability [1]. The Circle of Willis (CoW), a ring-like arterial network at the base of the brain, serves as a crucial anatomical structure responsible for maintaining balanced and consistent cerebral blood perfusion [2, 3]. However, only approximately 20% of people present a complete CoW [4-7]. The incomplete CoW, known as variation, is increasingly recognized as a predisposing factor for stroke, particularly in cases involving hemodynamic compromise or collateral insufficiency [8-10]. Therefore, accurate and timely assessment of the CoW is important for identifying individuals at elevated stroke risk and guiding appropriate clinical management.

Medical imaging on the cerebral vessels plays an essential role in risk assessment [11, 12]. Ultrasound-based transcranial color-coded Doppler (TCCD) has emerged as a promising tool for evaluating the CoW owing to its unique advantages; its free of ionizing radiation and contrast agent requirement make TCCD a safer option for repeated evaluations. Meanwhile, its relatively low cost, high portability, and ease of access render TCCD particularly suitable for use in resource-limited settings and routine screening scenarios. Noteworthily, TCCD can also provide real-time visualization of intracranial blood flow dynamics, allowing for the detection of stenosis, occlusion, and other hemodynamically significant abnormalities within the CoW vessels [13-15].

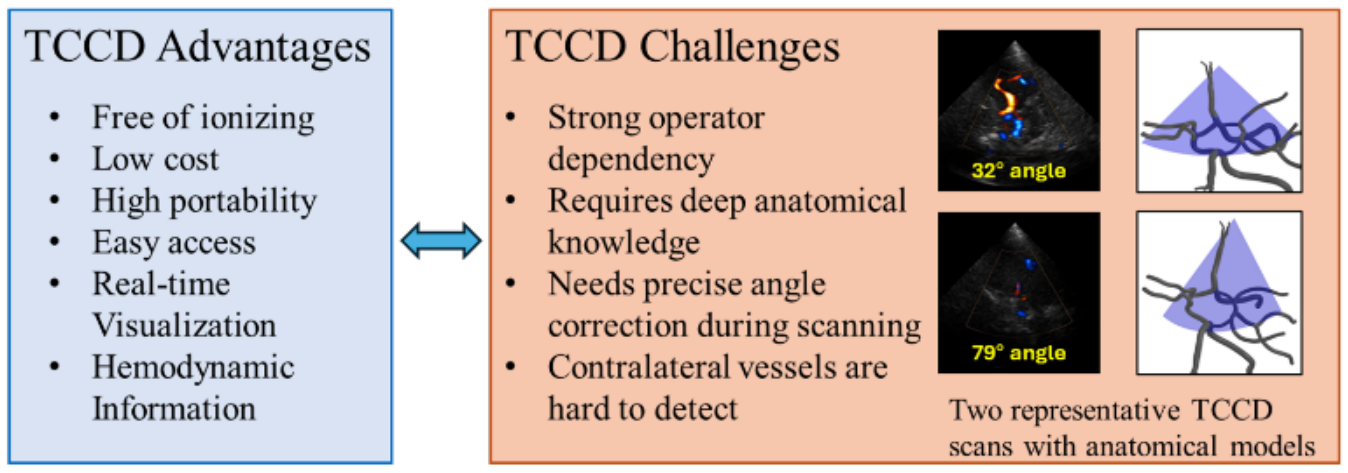

Fig. 1. Advantages and limitations of TCCD for CoW assessment, illustrated with representative scans and anatomical models. While optimal insonation angle (~32°) enables clear visualization of major arteries, steep angles (~79°) hinder detection of contralateral vessels. This highlights key limitations including operator dependency, anatomical variability, and the need for precise angle correction.

However, the adoption of TCCD for CoW assessment has been hindered by several limitations, primarily operator dependency. As illustrated in Fig. 1, clear visualization and accurate interpretation of TCCD images requires not only a deep understanding of cerebrovascular anatomy for vessel identification but also precise execution of angle correction during scanning. Variable vascular orientations and small vessels, especially at the contralateral side, could complicate the process, leading to inconsistent results among different operators [16, 17]. The reliability of TCCD as a diagnostic tool is compromised by this variability. As a result, the advantages of TCCD are still mostly limited to specialized facilities with knowledgeable sonographers, which restricts its wider use in primary care and rural health systems.

To overcome these challenges and enhance the consistency of TCCD-based CoW assessment, deep learning (DL) techniques have shown immense promise. Among the dominant approaches in medical image segmentation, U-Net variants and transformer-based architectures lead in accuracy and contextual modeling [18, 19]. However, their substantial computational demands often limit their use in real-time clinical settings [20]. In contrast, You Only Look Once (YOLO) frameworks are optimized for inference speed and efficiency [21], making them particularly well suited for the dynamic, frame-by-frame nature of TCCD imaging. Additionally, unlike conventional semantic segmentation, CoW analysis in TCCD involves multi-object segmentation, where several arterial structures may appear simultaneously within a single scan. YOLO's instance-level detection naturally fits this requirement. In the meantime, signal attenuation and anatomical variability make it difficult to accurately segment small, low-contrast contralateral arteries. Architectural enhancements that improve small object detection and segmentation are necessary to address this obstacle.

Despite the growing impact of DL across various imaging modalities, no existing study has yet applied these techniques to TCCD for cerebral vessel segmentation. A significant gap exists at the nexus of automated diagnostic support and cerebrovascular sonography This study aims to bridge the gap by leveraging the YOLO framework for CoW segmentation in TCCD. We further introduce an Attention-Augmented Wavelet YOLO (AAW-YOLO) architecture that combines lightweight attention modules with wavelet convolution, enhancing performance in anatomically complex regions—particularly where contralateral arteries appear small or indistinct within the sonographic field of view. Importantly, given the real-time nature of TCCD scanning and its critical reliance on dynamic image interpretation, evaluating the real-time feasibility of the model is essential. A solution that balances segmentation accuracy with computational efficiency could help translate AI-assisted TCCD from research into daily clinical workflows, especially in resource-limited or point-of-care settings.

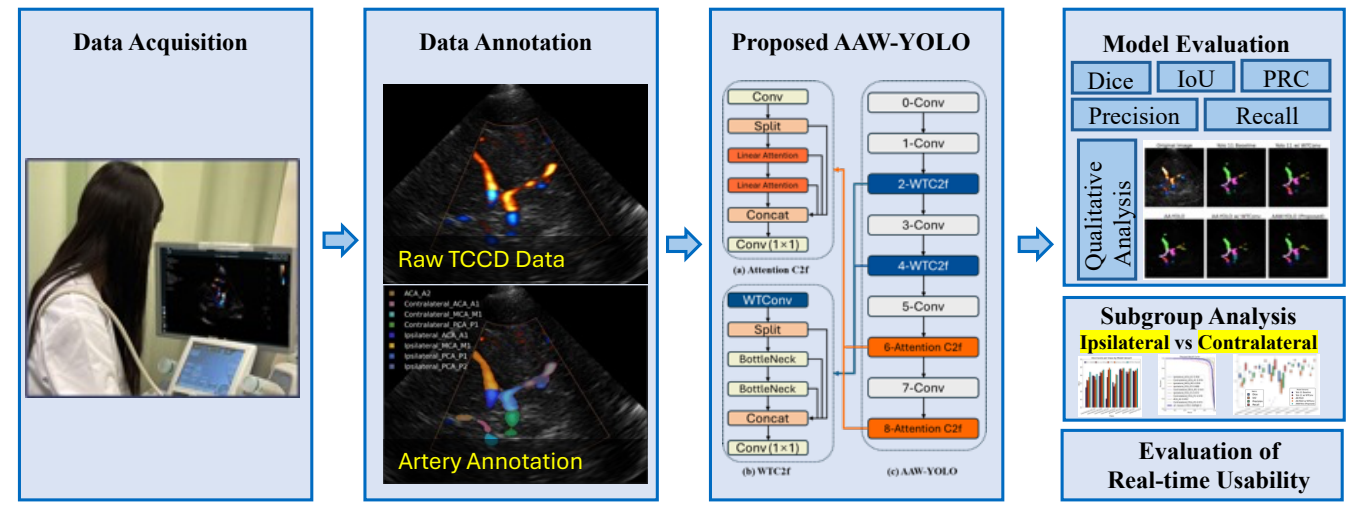

Fig. 2. Overview of the proposed AAW-YOLO workflow. The pipeline comprises data acquisition, data annotation, model training, model evaluation, subgroup analysis comparing ipsilateral vs contralateral arteries and evaluation of real-time usability.

On top of this, three specific objectives were defined in this study: (i) to assess the feasibility and segmentation performance of the YOLO framework for CoW analysis in TCCD imaging; (ii) to evaluate the impact of integrating wavelet-based and attention-enhanced modifications on segmenting contralateral arteries that exhibit greater anatomical complexity; and (iii) to explore the real-time applicability of the proposed AAW-YOLO, emphasizing computational efficiency and its potential for clinical deployment. The overall workflow of this study is shown in **Fig. 2**.

Key contributions of this study could be summarized as follows:

- To our knowledge, first-of-its-kind deep learning application for brain vessel segmentation on TCCD images.
- Developed AAW-YOLO, a lightweight architecture optimized for efficient cerebrovascular analysis.
- Provided subgroup analysis to assess consistency of model performance across ipsilateral and contralateral vessels.
- Offered a practical solution for real-time cerebrovascular assessment in clinical TCCD workflows.

## II. Methodology

### *A. Data Acquisition and Annotation*

TCCD data were prospectively collected from 15 subjects, comprising 29 videos recorded from both left and right insonation angles. Scanning was performed through the transtemporal acoustic window using ultrasound imaging equipment (Aixplorer model, SuperSonic Imagine, France) with a single-crystal phased array ultrasound probe (XP5-1, 96 transducer elements and 1-5 MHz of bandwidth). Ethical approval for this study was obtained from the Human Subject Ethics Sub-committee (HSESC) of the Hong Kong Polytechnic University (HSEARS20201119002). A total of 738 annotated frames containing 3,419 arterial instances were labeled to establish a high-quality dataset for model training and evaluation. All the annotations were performed following the well-established guidelines for cerebral vessel segmentation [22], ensuring consistency in identifying key arterial structures.

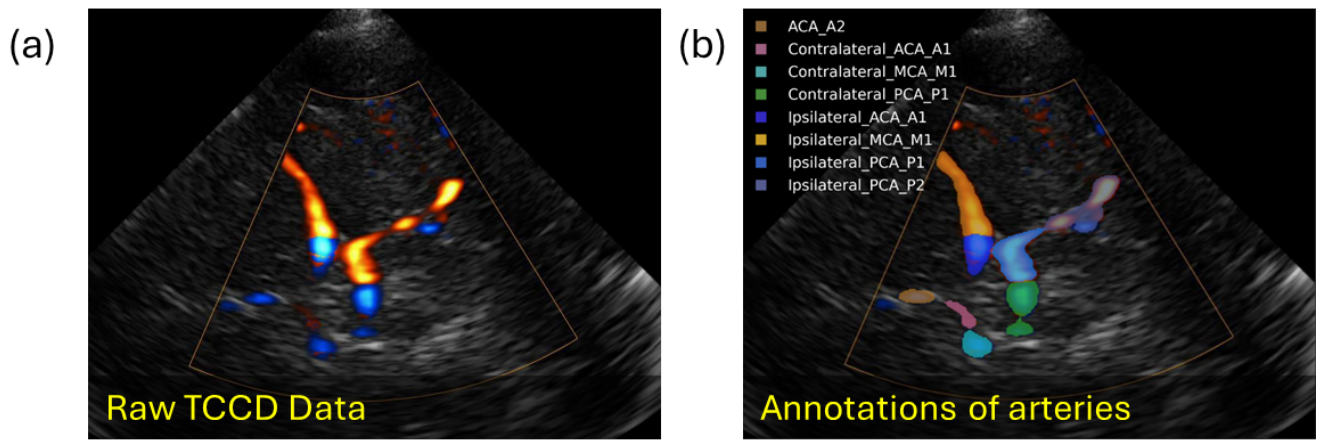

Fig. 3. Representative case of: (a) Raw TCCD Data. (b) Annotations of ipsilateral and contralateral cerebral arteries.

**Fig. 3** illustrates the annotation process. This study primarily focuses on annotations of the anterior cerebral artery (ACA), middle cerebral artery (MCA), and posterior cerebral artery (PCA). Arterial segments (A1, A2, M1, P1, P2) were defined according to standard cerebrovascular nomenclature [22, 23] with segment prefixes A, M, and P denote the corresponding artery group. Annotated artery segment instances include Ipsilateral_ACA_A1, Contralateral_ACA_A1, Ipsilateral_MCA_M1, Ipsilateral_PCA_P2, Contralateral_MCA_M1, Ipsilateral_PCA_P1, Contralateral_PCA_P1, Contralateral_PCA_P2 and ACA_A2. Due to the difficulty in distinguishing Ipsilateral vs. Contralateral ACA_A2, these segments were collectively labeled as ACA_A2, ensuring consistency in classification while preserving segmentation integrity. **Table I** presents the distribution of vessel segments in the annotated dataset, reported as instance counts and percentages. Ipsilateral arteries appear more frequently in the scans due to the attenuation of the ultrasound signal, with MCA particularly prominent because of its proximity to the transtemporal acoustic window. To balance the dataset, we intentionally increased the number of annotated cases for contralateral vessel instances.

TABLE I
DISTRIBUTION OF ANNOTATED ARTERIAL SEGMENTS

| Artery Segment | Instance Count | Percentage |
|---|---|---|
| ACA_A2 | 94 | 2.75% |
| Contralateral_ACA_A1 | 327 | 9.56% |
| Contralateral_MCA_M1 | 323 | 9.45% |
| Contralateral_PCA_P1 | 502 | 14.68% |
| Contralateral_PCA_P2 | 167 | 4.88% |
| Ipsilateral_ACA_A1 | 397 | 11.61% |
| Ipsilateral_MCA_M1 | 704 | 20.59% |
| Ipsilateral_PCA_P1 | 406 | 11.87% |
| Ipsilateral_PCA_P2 | 499 | 14.59% |

### *B. Selection of YOLO-11 as the Baseline Framework*

In this study, YOLO-11 [21] was chosen as the baseline and foundation framework. The selection was driven by its balance of efficiency, real-time processing capability, and multi-target segmentation proficiency [24, 25]. Its application in medical imaging and other biomedical fields has been widely validated, demonstrating high performance in precisely identifying complex anatomical structures while preserving computational viability [26-28]. Compared to U-Net-based [29-31] and transformer-based [32-34] architectures, YOLO-11 prioritizes speed and instance-level detection, making it particularly well-suited for dynamic medical imaging applications, such as TCCD vessel segmentation.

U-Net-based networks are well-regarded for precise pixel-wise segmentation, particularly in CT and MRI scans, but they often involve high computational costs and slower inference speeds [35, 36], making them less ideal for real-time TCCD analysis. Meanwhile, transformer-based networks, though powerful in capturing long-range dependencies, introduce significant overhead, making them computationally demanding for high-speed segmentation tasks [37, 38].

We acknowledge that UNext [39], an advanced U-Net derivative, emphasizes efficiency through parameter and FLOP optimization. However, while UNeXt demonstrates strong performance in single-lesion segmentation, its ability to handle multi-target scenarios remains uncertain. In tasks requiring the differentiation of overlapping or closely positioned vascular structures, more complex feature extraction mechanisms may be necessary to maintain segmentation precision [40]. YOLO is better suited for arterial differentiation in ultrasound imaging because of its instance-focused methodology.

By selecting YOLO-11 as the baseline framework, this study prioritized real-time efficiency and multi-target segmentation capabilities, ensuring a solid baseline for Doppler ultrasound vessel analysis. Its ability to distinguish overlapping arterial structures while maintaining fast inference speed makes it a well-suited foundation for advancing segmentation models in medical imaging.

### *C. Attention-Augmented Yolo (AA-YOLO)*

While transformer-based models are often computationally demanding, their predictive power remains strong. The integration of attention mechanisms within the YOLO framework has already been validated through YOLO-12[41] and is supported by a broader body of literature exploring attention-based enhancements in object detection [27, 42], demonstrating their effectiveness in enhancing feature representation. Our goal in this study was to make light use of this capability so that performance gains could be achieved without causing undue computational overhead. This was accomplished by proposing the Attention-C2F bottleneck block, which substitutes linear attention for bottleneck layers **(Fig. 4a)**.

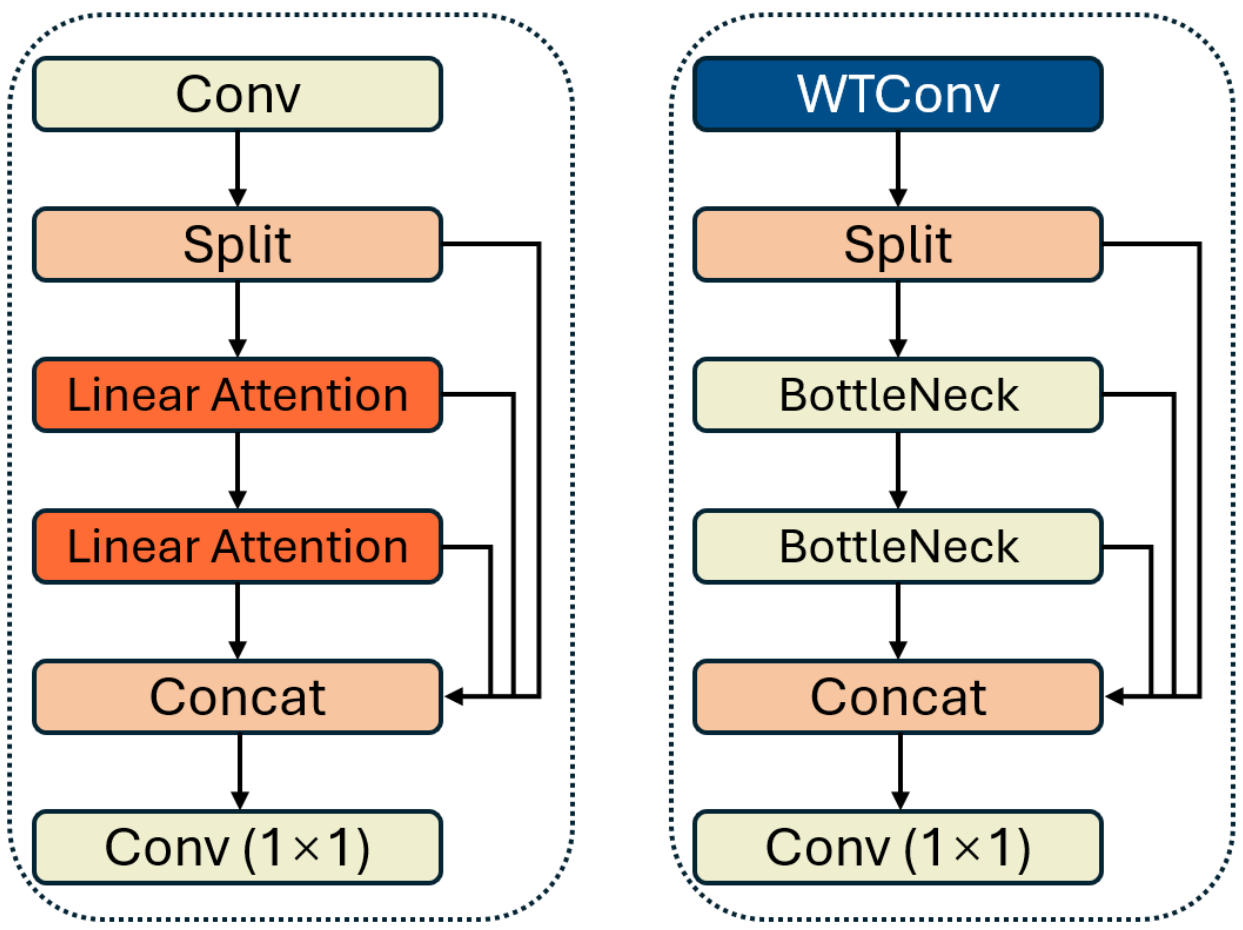

**(a) Attention C2f** **(b) WTC2f**

Fig. 4. Architectural diagrams of the proposed modules: (a) Attention C2f, and (b) WTC2f.

Bottleneck layers are commonly used to compress features [43, 44], which, while improving efficiency, risk the loss of fine vascular structures crucial for segmentation, especially in the case of small arterial structures, such as contralateral arteries in TCCD CoW images. These arteries, including contralateral ACA, MCA and PCA segments, are inherently more challenging to detect due to their lower contrast and proximity to other vascular structures. Replacing bottleneck layers with linear attention enhanced the focus on relevant arterial regions, enabling the segmentation model to prioritize small arteries over background artifacts.

Though linear attention may sacrifice part of global dependencies, this tradeoff aligns well with CoW artery segmentation, where precise localization of contralateral arteries is more critical than long-range spatial awareness. Furthermore, as shown in **Fig. 5c**, we replaced the original C3K2 blocks in YOLO-11 with the proposed attention-C2F block, strategically positioning them at higher levels of the backbone. At these higher stages, attention mechanisms better capture multi-scale vessel structures, allowing the network to refine arterial features while minimizing misclassification[45, 46]. This targeted enhancement ensures the model distinguishes contralateral vessels effectively, mitigating segmentation errors caused by overlapping tissues or insufficient contrast.

## D. Attention-Augmented Wavelet Yolo (AAW-YOLO)

Building upon AA-YOLO, this study incorporated Wavelet Convolution (WTConv) [47] to expand the model's receptive field, improving its ability to capture multi-scale vascular structures in CoW segmentation. WTConv leverages the Wavelet Transform as convolutional kernels, performing operations in the wavelet decomposing domain before reverting to the full image. This method effectively enlarges the receptive field, allowing the model to preserve fine vascular details while maintaining computational efficiency.

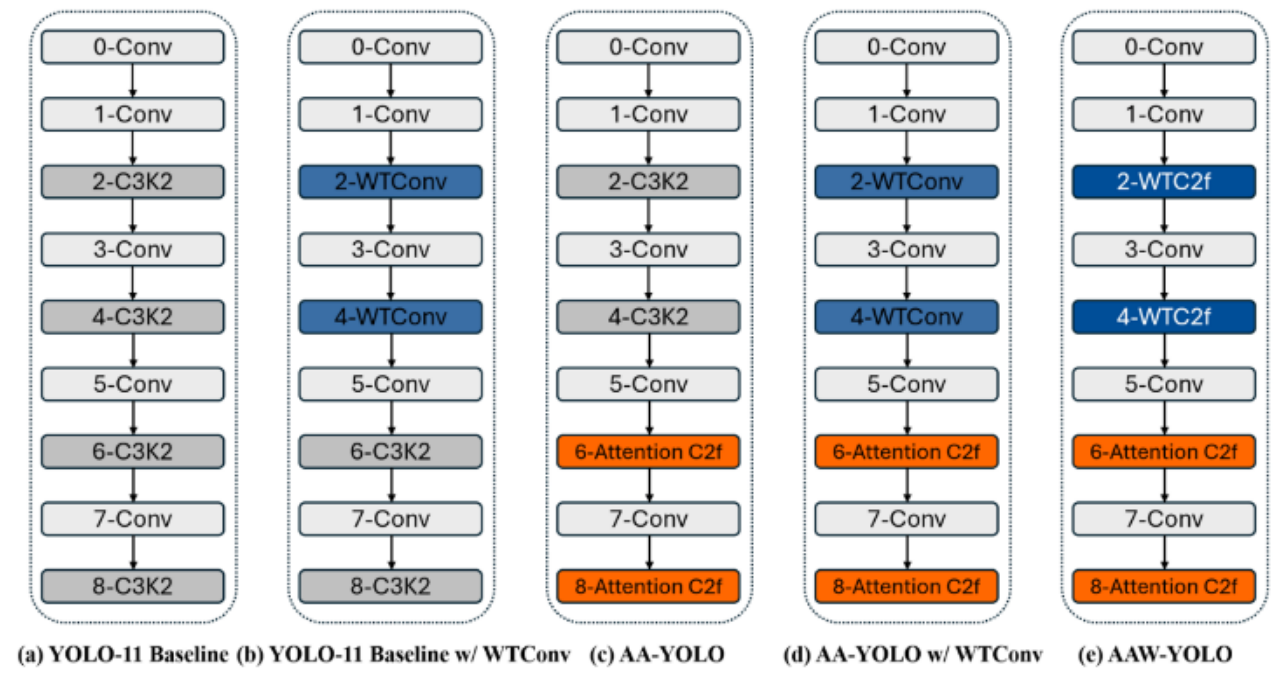

Fig. 5. Backbone structures of the proposed model variants: (a) YOLO-11 baseline, (b) YOLO-11 baseline with WTConv, (c) AA-YOLO, (d) AA-YOLO with WTConv, and (e) AAW-YOLO.

To further refine feature extraction, WTConv was integrated into the proposed Wavelet C2F bottleneck block, as shown in **Fig. 4b**, replacing the convolution head of the original C2F block with the WTConv block. In order to avoid distortions that could impair the performance of vascular segmentation, the bottleneck layers within the Wavelet C2F block were retained, striking a balance between efficiency and feature preservation [43]. This adjustment ensures that fine vascular structures are ca`ptured with greater clarity, particularly in small arterial regions, while maintaining the model's computational feasibility.

On top of this, we proposed AAW-YOLO, an extension of AA-YOLO that replaces the remaining C3K2 blocks in lower layers of the AA-YOLO backbone with the Wavelet C2F block, as shown in **Fig. 5e**. This modification was specifically designed to boost small object segmentation performance, ensuring that contralateral arteries and other fine vascular structures are effectively detected [21, 48].

## E. Experimental Settings and Evaluation Metrics

In this study, five model variants were proposed to systematically evaluate performance, including the YOLO-11 baseline, YOLO-11 with WTConv, AA-YOLO, AA-YOLO with WTConv, and AAW-YOLO, whose backbone structures are shown in **Fig. 5** for comparison. This arrangement was carefully designed to allow for comparative analysis, isolating the contributions of WTConv and attention mechanisms. The baseline YOLO-11 model serves as the reference, providing a foundation against which improvements can be assessed. By introducing WTConv separately into YOLO-11, its effect on receptive field expansion was evaluated independently. AA-YOLO, incorporating attention mechanisms, was tested to examine the improvements in feature refinement. The addition of WTConv to AA-YOLO allows for an assessment of their combined effects. Finally, AAW-YOLO, which integrates both enhancements into a unified framework, represents the most advanced model variant in this study.

Additionally, a subgroup analysis comparing ipsilateral and contralateral arteries was conducted to further refine the evaluation of model performance. This analysis would shed light on how well the model generalizes to different vascular structures, as contralateral arteries frequently pose more segmentation challenges because of their lower contrast and anatomical complexity. By assessing segmentation accuracy separately for ipsilateral and contralateral arteries, the study would shed some light on how each model variant handles small object segmentation, particularly in complex anatomical regions.

To assess the real-time usability of the proposed AAW-YOLO, key indicators—including parameter count, giga floating point operations (GFLOPs), inference speed, and corresponding frames per second (FPS)—were measured and compared across model variants. This evaluation was conducted on a single NVIDIA GeForce RTX 4070 Laptop GPU, ensuring real-time evaluation within a practical computational setup.

To quantitatively assess model performance, Dice coefficient, Intersection over Union (IoU), pixel-level Precision, pixel-level Recall, and mean Average Precision (mAP) were employed. These metrics provide a comprehensive evaluation of vascular segmentation performance and detection reliability, as demonstrated in a substantial body of studies focused specifically on AI-assisted medical image segmentation [18, 26, 49, 50].

Given the nature of TCCD imaging, where the background dominates over artery instances, conventional metrics such as specificity and accuracy tend to remain consistently high across different models. This is primarily because True Negatives (TN) are abundant—most pixels in the image belong to the background and are correctly classified as non-arterial regions. As a result, these metrics may fail to distinguish between model variants in a meaningful way, leading to their exclusion from the evaluation. Instead, we focused on metrics that directly assess segmentation fidelity and detection effectiveness within complex vascular structures. Below are the formulas for the evaluation metrics:

$$Dice = \frac{2 \times TP}{2 \times TP + FP + FN} \quad (1)$$

$$IoU = \frac{TP}{TP + FP + FN} \quad (2)$$

$$Precision = \frac{TP}{TP + FP} \quad (3)$$

$$Recall = \frac{TP}{TP + FN} \quad (4)$$

$$mAP = \frac{1}{n} \sum_{i=1}^{n} AP_i \quad (5)$$

In these equations, True Positives (TP) represent correctly identified arterial pixels, False Positives (FP) account for incorrectly labeled arterial pixels, and False Negatives (FN) represent missed arterial pixels. Additionally, $AP_i$ corresponds to the Average Precision for the i-th class, contributing to the overall mAP, which evaluates segmentation precision across multiple categories.

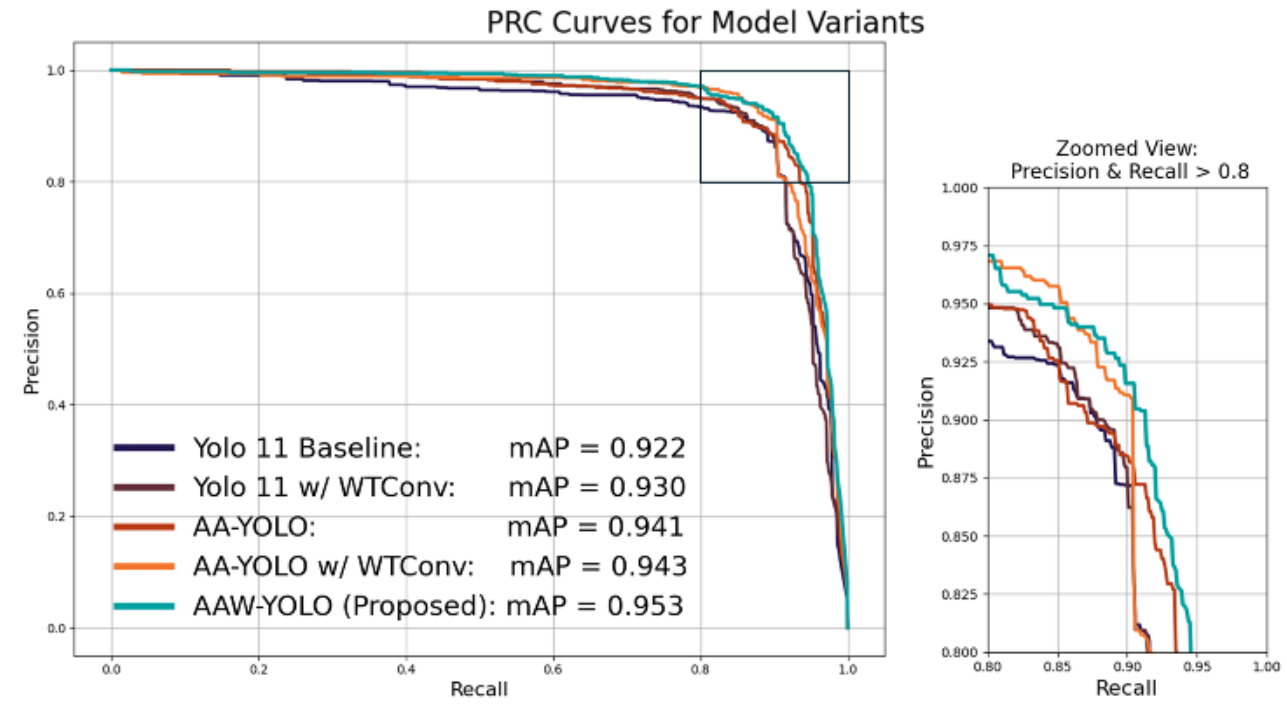

Fig. 6. PRC curves across model variants.

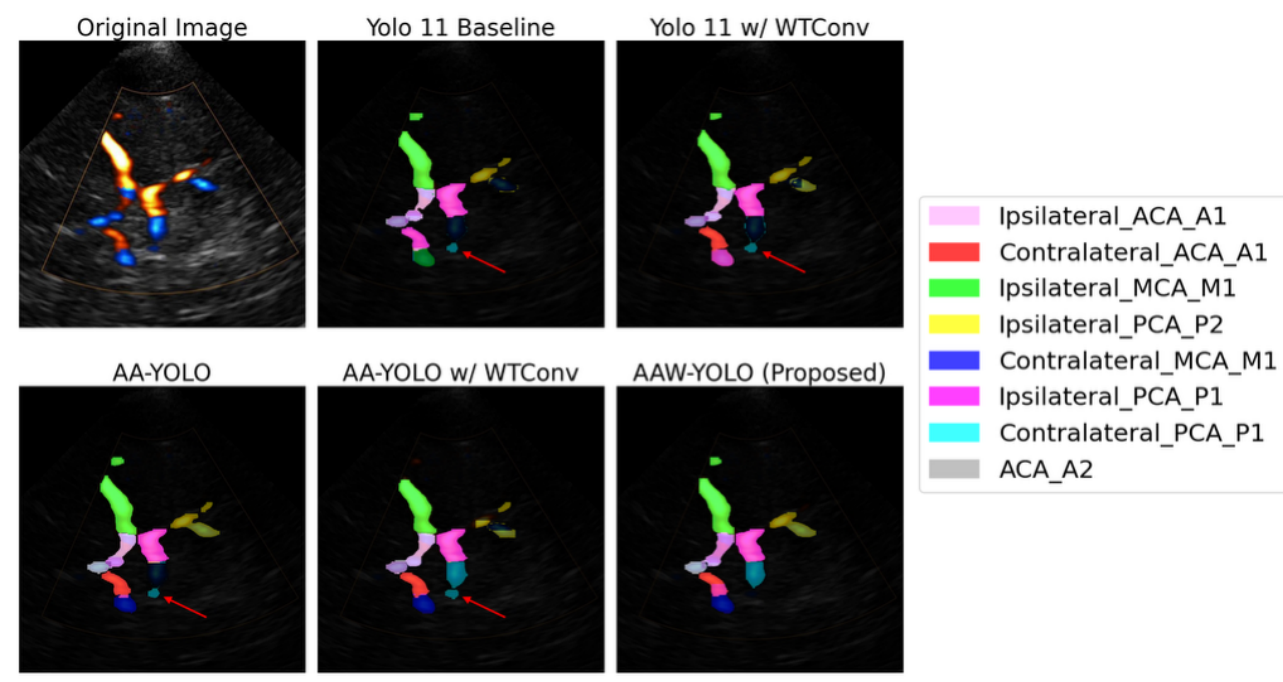

Fig. 7. Qualitative segmentation comparison across model variants.

TABLE II
SEGMENTATION PERFORMANCE ACROSS MODEL VARIANTS

| Model Variant | Dice | IoU | Precision | Recall | mAP |
|---|---|---|---|---|---|
| Yolo 11 Baseline | 0.868 | 0.773 | 0.841 | 0.908 | 0.922 |
| Yolo 11 w/ WTConv | 0.873 | 0.779 | 0.845 | 0.914 | 0.930 |
| AA-YOLO | 0.882 | 0.793 | 0.852 | 0.920 | 0.941 |
| AA-YOLO w/ WTConv | 0.883 | 0.796 | 0.857 | 0.919 | 0.943 |
| AAW-YOLO (Proposed) | **0.901** | **0.823** | **0.882** | **0.926** | **0.953** |

## III. RESULTS

### A. Segmentation Performance Overview

**Table II** presents the segmentation performance across different YOLO model variants, emphasizing the effectiveness of AAW-YOLO in achieving superior agreement with ground-truth annotations. The proposed model consistently outperformed its counterparts across key metrics, reporting average Dice coefficient (0.901), IoU (0.823), segmentation precision (0.882), segmentation recall (0.926), and overall mAP (0.953), suggesting that the integration of both attention mechanisms and wavelet processing yields a substantial improvement over baseline configurations, reinforcing its capability for enhanced segmentation accuracy and anatomical structure differentiation. This trend is visualized in **Fig. 6**, which presents the precision recall curves (PRC) across model variants.

A qualitative comparison, shown in **Fig. 7**, further illustrates these segmentation challenges by presenting outputs from multiple models applied to an ultrasound image. AAW-YOLO consistently provided a clearer segmentation with well-defined anatomical structures, whereas other models—including YOLO 11 Baseline, YOLO 11 w/ WT-Conv, and AA-YOLO—failed to delineate Contralateral_PCA_P1 (teal). Similarly, YOLO 11 Baseline, YOLO 11 w/ WTConv, and AA-YOLO w/ WTConv did not fully segment the complete Ipsilateral_PCA_P2 structure, reinforcing differences in segmentation reliability. YOLO 11 Baseline and YOLO 11 w/ WTConv also wrongly detected Contralateral_ACA_A1 and Contralateral_MCA_M1 as ipsilateral arteries.

Additionally, **Fig. 7** revealed an instance of incorrect segmentation, where a small teal dot appeared in the results of the first four models, as shown by the red arrows. This represents a false detection of Contralateral_PCA_P1, which should actually be an emerging Contralateral_PCA_P2. While the proposed model did not detect this emerging structure, it did not wrongly segment it either, avoiding unnecessary misclassification. Given that the vessel was in an early-stage appearance, its absence in segmentation remained understandable at this stage.

Together, these results demonstrated the proposed AAW-YOLO's superior performance, confirming its capacity to provide more accurate, dependable, and anatomically accurate vessel segmentations.

TABLE III
SUBGROUP SEGMENTATION PERFORMANCE OF IPSILATERAL AND CONTRALATERAL ARTERIES

| Model Variant | Dice | IoU | Precision | Recall |
|---|---|---|---|---|
| **Ipsilateral Subgroup** | | | | |
| **Yolo 11 Baseline** | 0.875 | 0.783 | 0.839 | 0.924 |
| **Yolo 11 w/ WTConv** | 0.877 | 0.78 | 0.852 | 0.912 |
| **AA- YOLO** | 0.872 | 0.78 | 0.832 | 0.926 |
| **AA-YOLO w/ WTConv** | 0.888 | 0.804 | 0.865 | 0.921 |
| **AAW-YOLO (Proposed)** | 0.903 | 0.829 | 0.882 | 0.932 |
| **Contralateral Subgroup** | | | | |
| **Yolo 11 Baseline** | 0.827 | 0.737 | 0.808 | 0.867 |
| **Yolo 11 w/ WTConv** | 0.832 | 0.735 | 0.81 | 0.883 |
| **AA- YOLO** | 0.84 | 0.747 | 0.796 | 0.904 |
| **AA-YOLO w/ WTConv** | 0.842 | 0.754 | 0.813 | 0.892 |
| **AAW-YOLO (Proposed)** | 0.877 | 0.798 | 0.85 | 0.922 |
| **Performance Difference (Ipsilateral - Contralateral)** | | | | |
| **Yolo 11 Baseline** | 0.048 | 0.046 | 0.031 | 0.057 |
| **Yolo 11 w/ WTConv** | 0.045 | 0.045 | 0.042 | 0.029 |
| **AA- YOLO** | 0.032 | 0.033 | 0.036 | 0.022 |
| **AA-YOLO w/ WTConv** | 0.046 | 0.05 | 0.052 | 0.029 |
| **AAW-YOLO (Proposed)** | **0.026** | **0.031** | 0.032 | **0.010** |

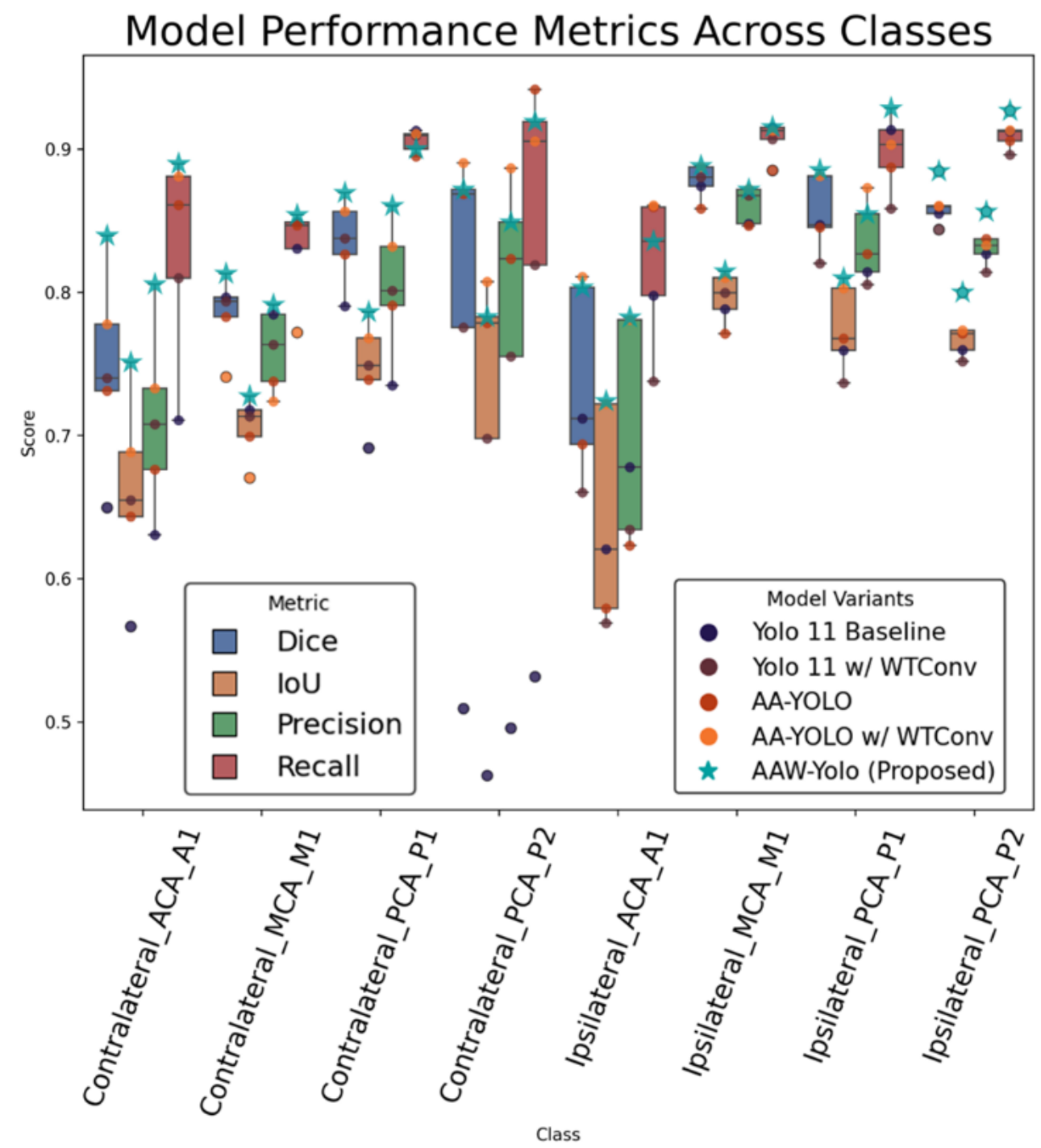

Fig. 8. Box-whisker plot of class-wise segmentation performance across model variants.

### *B. Subgroup Analysis: Ipsilateral vs. Contralateral Artery Segmentation*

**Table III** presents a thorough analysis of segmentation performance for ipsilateral and contralateral arteries. A clear pattern emerged that contralateral arteries reported lower segmentation performance compared to ipsilateral arteries. This discrepancy was reflected in Dice, IoU, precision, and recall metrics, with performance drops ranging from 0.022 to 0.057 across models, suggesting the increased complexity of distinguishing vessels on the contralateral side. As visualized in **Fig. 8**, a class-wise box-whisker plot further illustrates these segmentation differences across vessel subgroups. The plot highlighted how contralateral segmentation exhibits systematically lower accuracy, reinforcing the challenge of distinguishing vessels on the contralateral side.

However, our proposed AAW-YOLO achieved the smallest difference between ipsilateral and contralateral segmentation, with a Dice drop of 0.026 and an IoU drop of 0.031. Class-wise Dice and IoU scores are visualized in **Fig. 9** and **Fig. 10**, respectively.

Beyond maintaining the smallest performance drop, the proposed model also achieved the highest segmentation performance in both ipsilateral and contralateral artery subgroups, recording a Dice coefficient of 0.903 (ipsilateral) and 0.877 (contralateral), along with the highest IoU scores of 0.829 (ipsilateral) and 0.798 (contralateral).

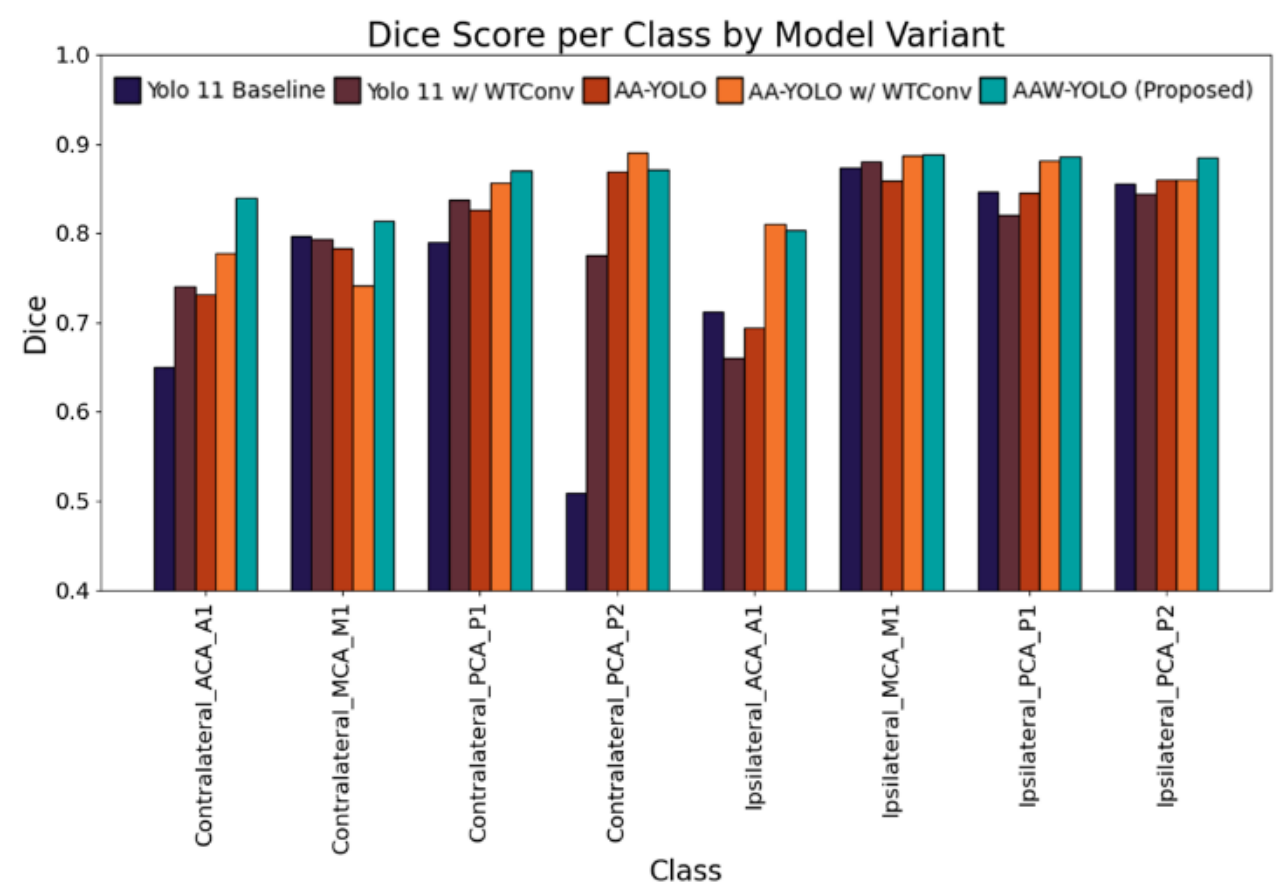

Fig. 9. Class-wise Dice scores.

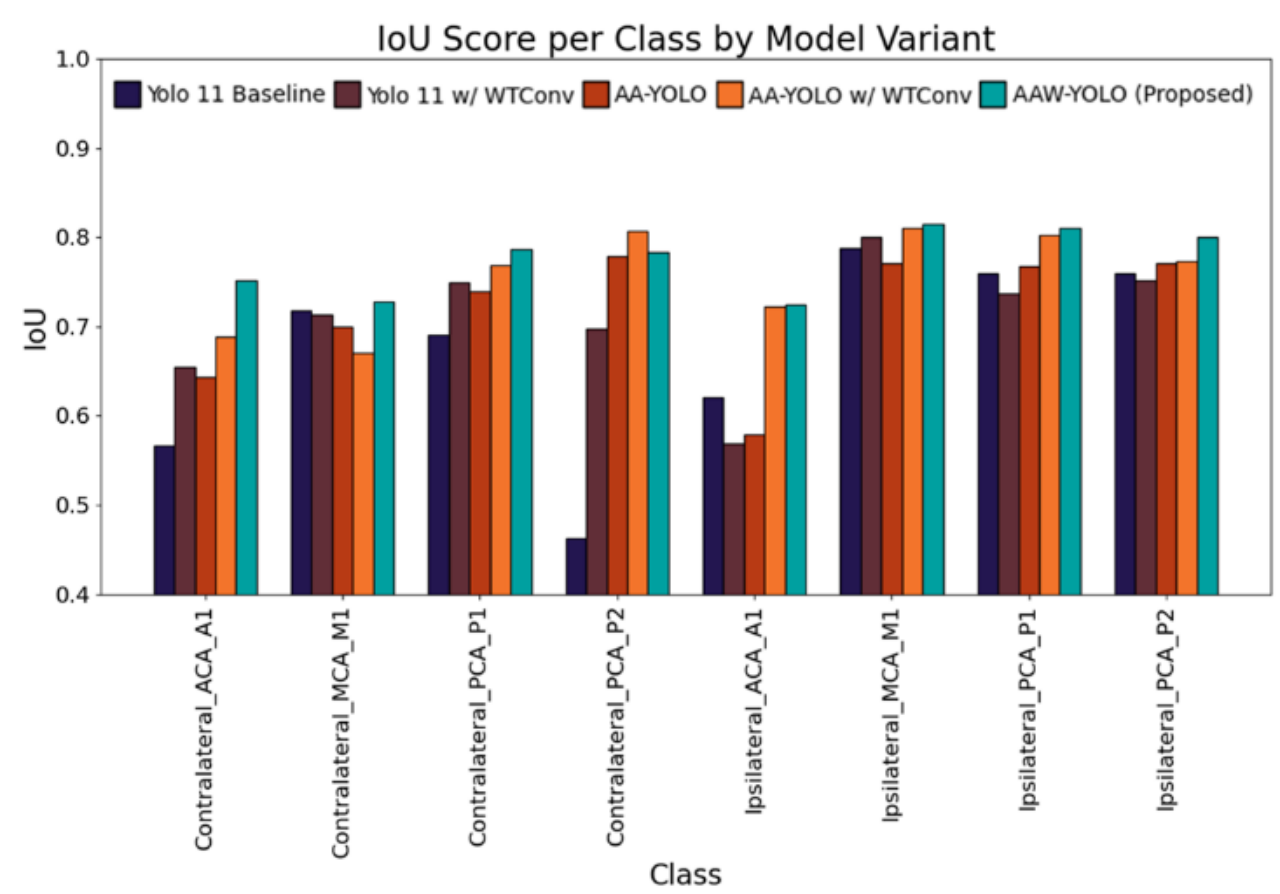

Fig. 10. Class-wise IoU scores.

**Fig. 11** further compares the variation of performance metrics across model variants, showing that the proposed AAW-YOLO consistently achieved the lowest coefficient of variation in Dice, IoU, precision, and recall.

Together, these results demonstrate that the proposed AAW-YOLO shows strong generalization to anatomical variability. It not only achieved the highest segmentation accuracy across ipsilateral and contralateral artery subgroups, but also maintained the most consistent performance across classes.

### C. Computational Efficiency

**Table IV** presents the efficiency metrics for various model variants, detailing parameter count, GFLOPS, per-frame inference speed, and corresponding FPS. All results were generated using a single NVIDIA GeForce RTX 4070 Laptop GPU, ensuring real-time evaluation within a practical computational setup, with performance metrics averaged over 1,000 test samples, maintaining manageable computational overhead. This result indicates a balanced trade-off between segmentation performance and computational efficiency, making it a viable solution for real-time applications of ultrasound-based vessel analysis.

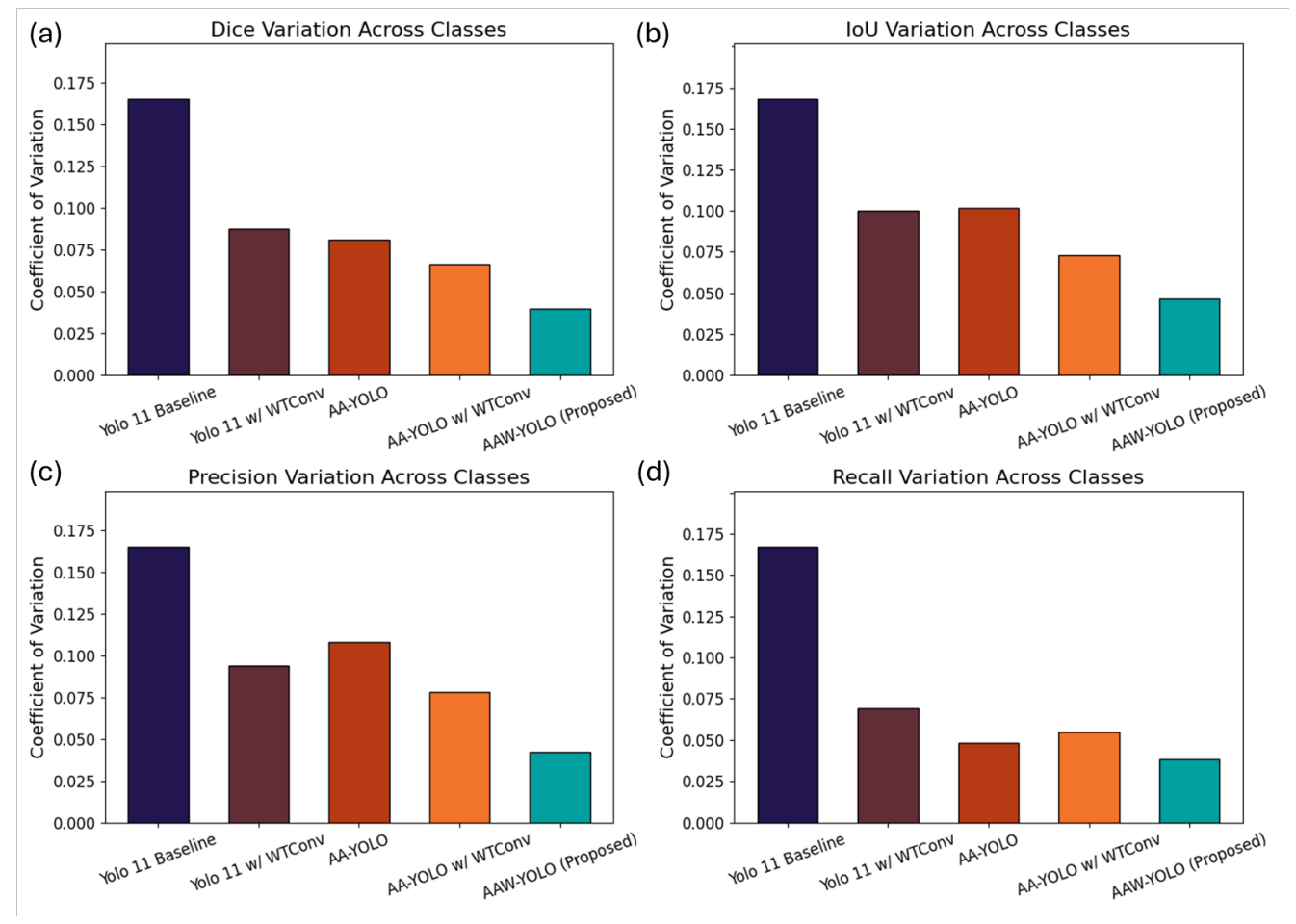

Fig. 11. Variation of performance metrics across model variants in Dice, IoU, precision and recall.

TABLE IV
COMPARISON OF COMPUTATIONAL EFFICIENCY ACROSS MODEL VARIANTS

| Model Variants | Parameters (M) | GFLOPS | Inference Speed (ms) | FPS |
|---|---|---|---|---|
| Yolo 11 Baseline | 2.844 | 10.4 | 8.363 | 119.574 |
| Yolo 11 w/ WTConv | 2.27 | 10 | 10.871 | 91.988 |
| AA-YOLO | 2.823 | 10.4 | 10.984 | 91.042 |
| AA-YOLO w/ WTConv | 2.711 | 9.3 | 13.703 | 72.977 |
| **AAW-YOLO (Proposed)** | **2.761** | **10.4** | **14.199** | **70.427** |

## IV. Discussions

### A. Key Insights from Segmentation Performance

This study validated the suitability of the YOLO framework for CoW segmentation in TCCD imaging. As shown in **Table II**, the proposed AAW-YOLO consistently outperformed alternative models across core metrics—Dice, IoU, Precision, Recall and mAP—demonstrating its reliability in achieving precise anatomical delineation. The integration of wavelet-based processing and attention mechanisms enhances low-level spatial sensitivity and contextual awareness, contributing to improved segmentation quality.

In alignment with our objective to enhance segmentation of challenging arterial regions, subgroup analysis reveals that contralateral arteries remain inherently difficult to segment, likely due to anatomical asymmetries and lower signal clarity. The proposed AAW-YOLO reduced this discrepancy by achieving a recall difference of merely 0.010 compared to its ipsilateral counterparts' performance, as shown in **Table III**, reflecting a substantial improvement in the model's ability to identify overlapping or low-contrast structures.

### B. Real-time Usability

As shown in **Table IV**, the proposed AAW-YOLO achieved an inference speed of 14.199 milliseconds per frame. While this is slower than the 8.363 milliseconds per frame recorded for the YOLOv11 baseline, it remains well within the real-time performance range required for typical clinical TCCD scanning. In typical clinical settings, TCCD scan operates at 20–30 FPS [51, 52]. While some high-end ultrasound systems can surpass 50 FPS, this often comes at the expense of resolution or penetration depth, limiting practical use [53]. Notably, the inference speed of our proposed AAW-YOLO (14.199 milliseconds per frame) translated to 70.427 FPS, which remains well above the ideal 20 FPS threshold for TCCD scanning and comfortably within the ultrasound system limit of 50 FPS. These results validated the proposed AAW-YOLO as a practical and efficient solution for real-time clinical settings.

### C. Study Strength

The proposed AAW-YOLO framework holds meaningful potential for advancing non-invasive cerebrovascular diagnostics in clinical and underserved settings. By achieving high segmentation performance at real-time inference speed, the proposed model reduces operator dependency, thereby enabling wider accessibility of TCCD technology and aiding early detection and monitoring of stroke-related pathologies, particularly in resource-limited or point-of-care environments.

To the best of our knowledge, this study presents the first AI-based approach for TCCD imaging that explicitly emphasizes the segmentation of contralateral vessels—an anatomically challenging task due to variable morphology, low signal clarity, and smaller vessel diameter. By integrating proposed WTC2f and Attention C2f blocks within the AAW-YOLO architecture, the model effectively enhances detection sensitivity in these difficult regions, addressing a long-standing technical limitation in cerebral vessel segmentation.

From a clinical standpoint, this is also the first work to assess both segmentation quality and real-time applicability of a DL model for CoW analysis in TCCD. Beyond comprehensive metric-based evaluation, the study incorporates practical deployment indicators such as FPS to establish real-world feasibility. Furthermore, the inclusion of a subgroup analysis provides insight into the model's robustness across cerebral arteries, particularly between ipsilateral and contralateral ones, highlighting its potential for consistent performance in dynamic real-world sonographic environments.

### D. Limitations and Future Directions

Several limitations should be acknowledged in this study. First, the current system focuses on frame-wise segmentation. As a result, it lacks awareness of sequential frames. Integrating a video-tracking practice into the model could potentially improve segmentation performance by enhancing logical consistency across consecutive frames. Second, this study was limited to unilateral TCCD analysis, which constrains the model's ability to leverage mirror-view anatomical information. Since arteries that appear contralateral in one scan are ipsilateral in the opposite insonation window, bilateral imaging would allow clearer visualization from both sides, potentially improving segmentation accuracy and robustness for small or ambiguously positioned vessels. Third, some vessels in the CoW—particularly the anterior and posterior communicating arteries (ACoA and PCoA)—remain challenging to visualize in TCCD due to acoustic attenuation from the skull. Additionally, in the case of the PCoA, its anatomical orientation often causes the direction of blood flow to align perpendicularly to the ultrasound beam (i.e., insonation angle near 90 degrees), at which point the Doppler signal cannot effectively detect flow, resulting in signal dropout or vessel invisibility [54]. As contrast-enhanced ultrasound can significantly improve vascular visibility, the absence of contrast agents in the current dataset may limit the model's ability to detect these small branches [55, 56]. Future studies incorporating contrast enhancement may enable better recognition of PCoA and ACoA, and offer a clearer assessment of the algorithm's capability to resolve subtle neurovascular anatomy. Forth, the dataset was obtained from a single-center cohort, which may introduce sampling bias or limit generalizability. Future validation across multiple centers with varied sonographic equipment and patient populations will be crucial for evaluating model generalizability and clinical readiness.

Future work should focus on these areas of expansion. Integrating temporal modeling—such as frame linkage, motion estimation, or trajectory-aware segmentation using architectures like XMem [57] or Space-Time Memory Networks (STM) [58]—could further improve segmentation performance and empower real-time sonographic guidance. In parallel, developing a pipeline capable of bilateral CoW analysis would enhance the system's diagnostic completeness by facilitating cross-window anatomical referencing, which would improve the visibility and segmentation performance of small, variably visualized arteries. Furthermore, incorporating contrast-enhanced TCCD could enable comprehensive evaluation of the model's capacity to segment vascular territories that remain poorly visualized under conventional Doppler conditions. Finally, multi-center studies involving diverse imaging settings, ultrasound vendors and populations are essential to confirm performance consistency and facilitate broad clinical adoption.

## V. Conclusion

This study marks the first implementation of deep learning methods for segmenting the CoW in TCCD imaging, offering a new pathway for automating cerebrovascular assessment with this modality. The proposed model—AAW-YOLO—achieved consistently outstanding performance across multiple evaluation metrics, reporting average Dice coefficient (0.901), IoU (0.823), precision (0.882), recall (0.926), and mean average precision (mAP) (0.953), while operating at an inference rate of 70.427 FPS which is well above the 20 FPS threshold typically required for clinical TCCD scanning, indicating its suitability for real-time applications. It also demonstrated improved performance in segmenting anatomically subtle structures, as reflected in a minimal recall gap of 0.010 between contralateral and ipsilateral arteries.

These results highlight the feasibility of combining high segmentation accuracy with fast computational performance in a lightweight model tailored for real-time sonographic analysis. While the study was limited to single-frame segmentation, unilateral views, and data from a single center, it provides a foundation for future advances in temporal modeling, bilateral analysis, and broader clinical validation. With further refinement, such approaches may contribute meaningfully to the development of AI-assisted tools for dynamic cerebrovascular screening and stroke prevention.